# Is Blockchain Hashing an Effective Method for Electronic Governance?

Oleksii Konashevych,[a,1] Marta Poblet[2]
[a]*Erasmus Mundus Joint International Doctoral Fellow in Law, Science, and Technology*
[b] *RMIT University, Graduate School of Business and Law*

**Abstract.** Governments across the world are testing different uses of the blockchain for the delivery of their public services. Blockchain hashing--or the insertion of data in the blockchain—is one of the potential applications of the blockchain in this space. With this method, users can apply special scripts to add their data to blockchain transactions, ensuring both immutability and publicity. Blockchain hashing also secures the integrity of the original data stored on central governmental databases. The paper starts by analysing possible scenarios of hashing on the blockchain and assesses in which cases it may work and in which it is less likely to add value to a public administration. Second, the paper also compares this method with traditional digital signatures using PKI (Public Key Infrastructure) and discusses standardisation in each domain. Third, it also addresses issues related with concepts such as "distributed ledger technology" and "permissioned blockchains." Finally, it raises the question of whether blockchain hashing is an effective solution for electronic governance, and concludes that its value is controversial, even if it is improved by PKI and other security measures. In this regard, we claim that governments need first to identify pain points in governance, and then consider the trade-offs of the blockchain as a potential solution versus other alternatives.

## 1. Introduction

Governments across the world are testing potential uses of the blockchain in their public sectors and, in particular, for property registries and notarisation of deeds. In 2016 Estonia launched a "notarisation on the blockchain" project with BitNation [1], although depriving such notarised acts of legal force. Some other countries have also started pilots: Honduras announced a blockchain-based real estate registry, but the project was eventually discontinued [2]; Chromaway—a Swedish start-up—announced in 2016 promising plans to upgrade the Swedish real estate registry by conducting records of deeds on the blockchain [3]. Yet, two years later the third phase of the pilot has been concluded but no results are yet available [4]. In the USA, the project Velox.re in Cook County (Chicago) tested a transaction outside of the real registry to imitate the use of the blockchain for deeds with real estate [5]. The clerk's office of the county issued a report [5] but the project never went ahead. Ubitquity.io announced the project Bitland to

---

implement a blockchain-based real estate registry in Ghana with no further continuity either [6].

Other pilots are currently work in progress. Ukraine and the Republic of Georgia announced their cooperation with Bitfury [7] to apply distributed ledger technology and blockchain to their cadastral registries.

Different organisations in the EU and UK have recently released reports on the use of the blockchain [8] [9] [10]. These reports generally express positive views about the impact of the technology and its value in the development of the informational society. Yet, they are much less specific about the design of blockchain-based e-government systems and how to implement blockchains in particular areas.

In this paper we raise some caveats about the use of the blockchain for public—state-owned—registries. We contend that the present lack of standards and regulatory frameworks makes its adoption contentious, as in some cases it could undermine e-government services and public interest in general. In this context, we analyse the potential of blockchain hashing in the e-government space and compare this method with the more traditional ways of digital signatures using standardised PKIs (Public Key Infrastructures) such as eIDAS in the European Union [11]. We conclude that governments should identify and consider pain points in the administrative processes before making decisions that involve blockchain adoption.

## 2. How does hashing on the blockchain work?

Broadly, hashing on the blockchain refers to inserting a hash sum in a blockchain transaction. Insertion implies that data is "published in the ledger and cannot be censored or retracted and will be permanently available to the world" [12]. The paper [12] offers a comprehensive analysis of different methods of data insertion in Bitcoin.

The method of hashing is recognised as a way of securing public data on the blockchain. Each entry in the central database of the public registry is hashed and casted to the blockchain. This concept is considered by some governments as a way of the use of the blockchain. An example of this is the project by Bitfury in Ukraine, developed in partnership with Transparency International (TI) and the Ukrainian government. The project, launched in 2017, applies Bitfury's DLT "Exonum" to hash records of the geocadastral registry [7]. Authorised nodes that are controlled by the government cast hashes to the Exonum-based ledger. The hash of the current state of the ledger is periodically anchored on the public blockchain. Initially, this public blockchain was Bitcoin, now it is Emercoin [13]. TI keeps another node, which plays the role of the observer and has permission to read the ledger only. As we see, Exonum is an example of a private permissioned DLT. Technical details can be found at [13].

## 3. Digital signatures vs hashing on the blockchain

Asymmetric cryptography consists of a pair of keys: public and private. The public key is a code string that uniquely identifies a certain individual or company. While the public key can be shared with anyone, the private key must be kept secure and undisclosed [14].

In essence, digital signatures and blockchain hashing do the same thing. Both methods leverage cryptographic algorithms to detect the forgery of an original entry. If any bit of information is altered, the hashes will not match. Since the signing of a blockchain transaction is based on asymmetric cryptography [15] the cryptographic signing of data is not different from signing blockchain transactions. The main difference between the two methods is that hashing on the blockchain adds another layer of data. With digital signatures, users insert their data as an INPUT to the cryptographic function; with hashing on the blockchain, users insert payment data along with the required data.

Digital signatures have not been used for public purposes in this basic form. For this to happen, the system needs to be supported by a Public Key Infrastructure (PKI) which includes a set of standards. PKI consists of technologies, procedures and actors that enable deployment of public-key cryptography-based security services [16]. Within PKI, one provider—known as Certificate Authority (CA) or Trust Service Provider (TSP)—is responsible for the provision of identity services, while other providers—Timestamp Authorities (TSA)—authenticate time and date information. Blockchain hashing, in contrast, does not require a centralised TSA since all blocks are chronologically stored. Timestamps are embedded in the blocks and remain immutable.

At present, PKI benefits from a complete infrastructure with regulations, standards, and procedures. For example, a roadmap of standards in the EU is described in [17]. In contrast, there is no standardised protocol to manage access in permissioned blockchains. There are no procedures for authorising nodes and operators (clerks) either. Clerks use private keys, but there is no standardised protocol if these keys are compromised. More generally, there is no standard for DLTs.

DLT and the blockchain, in summary, are not fully equipped yet. Without regulations, standards, procedures, and certifications, hashing on the blockchain can only work in limited environments, under supervision and, presumably, for research purposes. This is definitely not a scalable approach for e-government at this point in time.

**4. How hashing on the blockchain should work?**

There are at least three issues that must be addressed to leverage hashing and improve the security of centralised public registries: (i) identification; (ii) bi-directional relations of entries in the database and hashes on the blockchain; and (iii) standardisation.

*4.1. Identification*

The blockchain was initially conceived to preserve the anonymity (or, at least, the pseudonymity) of both nodes owners ("miners") and owners of blockchain addresses ("users"). Therefore, blockchains and DLTs must be supplemented with overlaid solutions for identification, authorisation and authentication, with the component of trust services (similar to eIDAS) where necessary.

In a permissioned DLT, the administrator will keep the key to the system and grant permissions (therefore, standards and procedures for managing keys and accesses must be applied). However, if hashing takes place on a public blockchain no one will know who did that—an authorised officer or an attacker—and therefore addresses will need to be identified, or the inserted data will have to contain an identifiable digital signature. Moreover, if the private key to the blockchain address is compromised or stolen, a procedure for a stop list where such an address is added to the special database and any further action from this address is considered invalid will be required. Again, we come back to the need for PKI standards based on proved principles.

*4.2. Bi-directional relations between databases*

The probable attack scenario here is that the record on the central database is changed or replaced with a new one. If there is no reverse relation with the hash stored on the ledger, a new hash can be created for the corrupted record and published in the distributed ledger, and so presented as a correct one. Apparently, such use of the blockchain does not add any value in terms of the security.

The issue is that hashing does not protect the data itself from being changed and deleted. It just helps to reveal the forgery if the user still keeps in their hands the original record. If the database is closed and centrally controlled—which is how databases work in the public administration—such manipulation is possible.

Traditionally, in government registries this issue is addressed by deploying a multilayered system of security measures, logs, and access procedures. One of the possible ways to use the blockchain is to build a central registry in the style of "chain of blocks" (chain of records, actually) similar to the blockchain, either public or private, or just to move the registry to the blockchain. Therefore, the use of the blockchain for public administration implies both the transfer of the central base to the blockchain—and not just hashing—and the use of PKI for identification.

*4.3. Permissioned blockchains, standardisation, and public policy*

The current hype about the blockchain may lead to some confusion when it comes to adoption by governments. The alleged achievements in the blockchainisation of e-government services tend to refer to "permissioned blockchains." Yet, most permissioned blockchains are architecturally centralised and have a hierarchical system of nodes. In this context, using the notion of "blockchain" or "permissioned blockchain" to refer to a centralised system can be misleading. Arguably, what qualifies as a blockchain is still under discussion: Original Bitcoin-like networks only? Decentralised systems based on other types of consensus mechanisms? A mix of them? What properties does a network have to exhibit to qualify as a blockchain? Ultimately, governments need to rely on shared conceptual frameworks and standards to make the appropriate technology choices. Otherwise, we may end up with scenarios where different government departments use different types of conflicting blockchains, or apply blockchain protocols with only a few nodes (unable to maintain the required level of security of the network), or use DLTs that are not blockchains at all.

Another risk associated with the lack of standardisation is that governments must keep the list of public blockchains whose technologies are proven trustworthy. Yet, such lists of "trusted" blockchains can lead to discrimination, arbitrarily excluding potentially appropriate networks and technologies. Standardisation is a better way to ensure fair competition and stable development.

Public policy and clear roadmaps are needed to avoid voluntarism and data breaches. Yet, a consistent blockchain public policy should also openly acknowledge the role of cryptocurrencies. Cryptocurrencies are the blood of the system, a key mechanism and incentive to maintain large networks sustainable. The alternative to cryptocurrencies are governments creating and maintaining decentralised network infrastructures. Ultimately, this leads to architectural decentralisation by political centralisation. A difficult balance to strike.

## 5. Conclusion

The standardisation of the blockchain domain is still in the early stages [18]. For this reason, hashing on the blockchain may be premature for public registries. Rather, the alternative option of PKI supported with standards and regulation is a more plausible solution at the present stage. Why, and what for, should we use the blockchain in e-government? Does the first generation of blockchains improve the security standards of existing centralised databases? These questions remain open. Our claim is that governments need first to identify pain points in governance, and then consider the trade-offs of the blockchain as a potential solution versus other alternatives

Arguably, governments will need to rethink the nature of public databases and perhaps look in the direction of the new generation of smart contracts and decentralised applications (DApps) (Blockchain 2.0). Smart contracts [19] and DApps can leverage many more functionalities than blockchain hashing, which is primitive and does not unleash the full potential of the blockchain. They could also help to reduce human-generated mistakes, bureaucracy, and costs for the public administration. Yet, smart contracts and DApps will equally require not just consistent regulatory frameworks, but broader organizational change in public administration and governance.